# Structures generated in a multi-agent system performing information fusion in peer-to-peer resource-constrained networks

Horacio Paggi · Juan A. Lara · Javier Soriano

**Abstract** There has recently been a major advance with respect to how information fusion is performed. Information fusion has gone from being conceived as a purely hierarchical procedure, as is the case of traditional military applications, to now being regarded collaboratively, as holonic fusion, which is better suited for civil applications and edge organizations. The above paradigm shift is being boosted as information fusion gains ground in different non-military areas, and human-computer and machine-machine communications, where holarchies, which are more flexible structures than ordinary, static hierarchies, become more widespread. This paper focuses on showing how holonic structures tend to be generated when there are constraints on resources (energy, available messages, time, etc.) for interactions based on a set of fully intercommunicating elements (peers) whose components fuse information as a means of optimizing the impact of vagueness and uncertainty present message exchanges. Holon formation is studied generically based on a multiagent system model, and an example of its possible operation is shown. Holonic structures have a series of advantages, such as adaptability, to sudden changes in the environment or its composition, are somewhat autonomous and are capable of cooperating in order to achieve a common goal. This can be useful when

Horacio Paggi
ETS de Ingenieros Informáticos. Universidad Politécnica de Madrid.
Tel.: +589-99165376
E-mail: horacio.paggi.straneo@alumnos.upm.es

*Present address:* ETS de Ingenieros Informáticos. Universidad Politécnica de Madrid. Campus de Montegancedo, 28660 Boadilla del Monte (Madrid), Spain

Juan A. Lara
Universidad a Distancia de Madrid (UDIMA). Carretera de La Coruña, KM.38,500 Vía de Servicio, nº 15 28400 Collado Villalba. Madrid, Spain

Javier Soriano
ETS de Ingenieros Informáticos. Universidad Politécnica de Madrid. Campus de Montegancedo, 28660 Boadilla del Monte (Madrid), Spain.

the shortage of resources prevents communications or when the system components start to fail.

# 1 Introduction

As information fusion gains ground in new areas (because of either the increasingly heterogeneous information sources or the diversity and amount of fused data) [8,54,55], the holarchy, which is a more flexible organization or structure for the information fusion than the hierarchy (originally used in military applications) because it exploits the autonomy of functional units [24], appears to be a better option. Holarchies are hierarchical organizations formed by holons, entities that have characteristics that provide a series of advantages, such as flexibility to adapt to changes in the environment and make rational use of resources: they are semi-autonomous and have a recursively self-similar structure like fractals. The formation of such holarchies has been little explored. In this respect, this article focuses on how holonic structures tend to be generated when the possible interactions in a system whose components perform information fusion to optimize the quality of the information gained and all interactions are possible (that is, peer-to-peer systems where there is a communication channel between any pair of system elements) are constrained (in the sense that there is something preventing communication from taking place even though there is communication channel between pairs). The aim is to minimize the failings associated with low quality (such as indeterminacy, basically caused by vagueness and uncertainty in received messages) to provide more certainty with respect to the content (semantics) of the message. This can be very important, in decision making for example. In this respect, the proposed model could, therefore, be said to be bio-inspired, especially taking into account research by [17] about how people make sense of the information that they receive ("sensemaking"). To illustrate holon formation, we use a system model that switches from operation in "unrestricted" to "intelligent" mode when its agents detect certain conditions, such as too many unanswered messages, as described by [40]. Systems that use the designed model self-organizing and form specific structures. Due to the strengths of the holonic structures (which will be addressed throughout this article), a system based on this model is applicable in open contexts made up of components that can fail or stop responding at any time and that require the highest possible quality response using the available resources.

While in [40] a model based on soft sensors using IF which allows to handle vague and uncertain communications is proposed by the authors, in this paper they continue the study of the properties of that model such as the emergent organizational structures. Additionally, this work exemplifies the holons formation in resource-constrained networks in decision making as a way of providing good options with acceptable costs. According to [37], they output very efficient algorithms working with local information to search for locally optimal solutions (that is, agent information and information provided by the agent environment). The idea here is to show how the search for efficiency within a P2P network yields certain specific structures.

This article is structured as follows. Section 2 introduces basic concepts concerning holons and self-organization and outlines a formal approach to holonic multiagent systems. Section 3 states the central problem of this research. Section 4 then briefly describes the operation of a system that implements this model and pinpoints the characteristics that it has that favor self-organization. Section 5 shows why a system of this type produces holons, and Section 6 illustrates an example of system operation based on this model. Finally, Section 7 outlines the conclusions and future work.

## 2 Related work

2.1 Fundamentals of holonic structures

*2.1.1 Holon*

In his book "The ghost in the machine" (1967), Arthur Koestler introduced the concept of holon to refer to entities that can be simultaneously regarded as a whole and a part: "which behave partly as wholes or wholly as parts, according to the way you look at them". Such entities could be living beings or social organizations [27]. Holarchies are dynamic hierarchies composed of holons that have the following characteristics: they make efficient use of their resources, are robust to changes in their environment, and are flexible and adaptable [9]. Holarchies can be said to guarantee global hierarchical control stability, predictability and optimization [52]. Holarchies are neither a static hierarchy nor a centralized structure, although they do have properties of both and are consistent with edge structures defined by [3]. The meaning of "edge" in this context is related with decentralized command and control structures, where each element can access a Global Information Grid, as Alberts writes [3].

Broadly speaking, holons cooperate benevolently to achieve a goal; it is not a blind benevolence where every holon spends all its time in new cooperations with the others. On the contrary, it constantly reconsiders its commitments and duties and only rejects collaboration when the required actions are impossible or highly unfavorable for the process. In this sense, holons can be called semi-autonomous. From the multiagent systems (MAS) standpoint, there are two forms of cooperation: explicit (the commitment is established through communication) and implicit (holons are designed such that goal-driven behavior emerges from subholon (that is, agent) behavior. The industry-level application of holons has been studied at length (see, for example, [29,30] [9, 32] [45]. Turnbull [49] separates strong holons from weak holons. Strong holons can exist autonomously, whereas weak holons cannot exist without the rest of the holarchy. In this paper, we consider only strong holons. Examples of natural holons are a cell, a human liver or the solar system; an semi-autonomous multiagent system, a workgroup or a (holonically organized) factory could be examples of artificial holons. Holons can be regarded as embodied abstract agents (that is, agents that have a physical part interacting with their environment). An abstract agent is an agent which can be considered as formed by other agents. Giret and Botti define: An abstract agent is a software system with a unique entity, which is located in an environment, which, as a whole, perceives its environment (environment sensitive inputs). From these perceptions, it determines and executes actions in an autonomous and flexible way - reactive and proactive. From a structural point of view, an abstract agent can be an agent (atomic entity); or it can be a multi-agent system (with a unique entity) made up of abstract agents that are not necessarily homogeneous. A multi-agent system is made up two or more abstract agents that interact to solve problems that are beyond the individual capabilities and individual knowledge of each abstract agent [10].

The most widespread holon structure is Fischer et al.'s [18] head-body holon, where the head performs all the holon's external communication and coordinates the body's activities. The body performs all the other holon functions (see Figure 1).

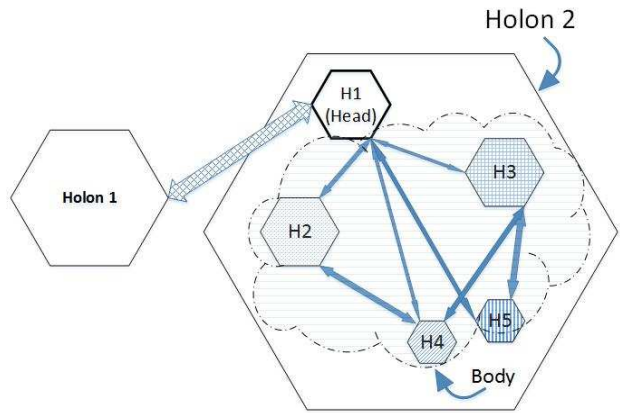

**Fig. 1** Holon head-body diagram (arrows represent communications, and H1, H2... denote holons)

### 2.1.2 Advantages

The adaptability to changes in the environment is typical of holons. This is patent when the holons are living beings.

Control is distributed such that control-related activities do not all fall to the delegator but are partially fulfilled by the delegate. This leads to fewer communications [43].

When rationality is limited (the information and communications processing capability is bounded) and the speed of adaptation to the environment is an important feature, systems composed of 'large' blocks perform better than systems composed of 'small' blocks [51], hence the importance of holons (where blocks mean a set of agents acting in coordination). Simon used the clockmaker parable to demonstrate this point [44].

2.2 Formalization

There have not been many attempts to formalize holon generation within a system. [47] gives a formal definition of holon, including the exchange between the whole and parts, and accounts for system data and knowledge. [19] and [22] penned more general papers. [22] shows:

- that any multiagent system containing $n$ agents can be mapped by means of an isomorphism to a system where only one agent is explicitly represented and the others are integrated within the system environment
- how to build an isomorphic system of a given MAS in which a group of agents are linked to a holon.

They also demonstrate that the holons within a MAS with the composition operation form an Abelian monoid (that is, an algebraic structure whose operation is associative, commutative and there is a neuter (the holon does not do anything))).

The above is described formally below [22]:

> **Definition 1. Multiagent system** (taken from [22])
> A multiagent system is a tuple $(\mathcal{A}, \varepsilon, \Pi, \Delta)$ where $\mathcal{A} = \{\alpha_1, \alpha_2, ... \alpha_n\}$ is the set of all agents. Each agent $\alpha_i$ is represented by a tuple $(S_i, P_i, A_i, \phi_i)$, where
> 
> - $S_i$ is the set of all possible agent states,
> - $P_i$ is the set of possible perceptions
> - $A_i$ are the actions that it can perform
> - $\phi_i$ is the function that describes agent activity $\phi_i : S_i \times P_i \to S_i \times A_i$. $\phi(s, p) = (\phi^1(s, p), \phi^2(p, s))$, that is, $\phi^1$ denotes the agent state and $\phi^2$ stands for the executed action
> - $\varepsilon$ is the set of possible states within the system environment
> - $\Pi$ is the perception function $\Pi : \varepsilon \to P_1 \times ... \times P_n$
> - $\Delta$ is the state of the system environment $\Delta : \varepsilon \times A_1 \times ... \times A_n \to \varepsilon$

> **Definition 2. System state change function**
> This is the function $\bar{\Delta} : \varepsilon \times S_1 \times ... \times S_n \to \varepsilon \times S_1 \times ... \times S_n$ o sea $\bar{\Delta}(e, s_1, ..., s_n) = (e', s'_1, ..., s'_n)$

> ***Definition 3. Isomorphic systems***
> Two MAS $(\mathcal{A}, \varepsilon, \Pi, \Delta)$ and $(\mathcal{A}', \varepsilon', \Pi', \Delta')$ are isomorphic if there is a bijection $\Psi : \varepsilon \times S_1 \times ... \times S_n \to \varepsilon' \times S'_1 \times ... \times S'_m$ such that, for any $(e, s_1, ... s_n) \in \varepsilon \times S_1 \times ... \times S_n$,
> $$\bar{\Delta}'(\Psi(e, s_1, ..., s_n)) = \Psi(\bar{\Delta}(e, s_1, ..., s_n))$$

The respective theorem is:

> ***Theorem***. An isomorphic multiagent system $(\{\alpha', \alpha_{k+1}, ..., \alpha_n\}, \varepsilon, \Pi', \Delta')$ can be built for any multiagent system $(\{\alpha_1, ..., \alpha_n\}, \varepsilon, \Pi, \Delta)$ and $k \leq n$, where $\alpha'$ is the holon formed based on $\alpha_1, ..., \alpha_k$.

Gerber et al. proof the existence of $\alpha'$ by constructing $\alpha' = (S', P', A', \phi')$, such that, by definition:

- $S' = (S_1 \times ... \times S_k)$
- $P' = (P_1 \times ... \times P_k)$
- $A' = (A_1 \times ... \times A_k)$
- $\phi'(s', p') = ((\phi_1^1(s_1, p_1), ... \phi_k^1(s_k, p_k)), (\phi_1^2(s_1, p_1), ... \phi_k^2(s_k, p_k)))$
- $\Pi' : \varepsilon \to (P', P_{k+1}, ... P_n)$ such that $\Pi'(e) = ((\Pi_1(e), ..., \Pi_k(e)), \Pi_{k+1}(e), ... \Pi_n(e))$ for any $e$
- $\Delta' : \varepsilon \times A' \times A_{k+1} \times ... A_n$ such that $\Delta'(e, a', a_{k+1}, ... a_n) = \Delta(e, a_1, ... a_n)$

A hypothesis implied by this theorem is that all the agents can communicate and collaborate with all the others, as, otherwise, it could be impossible to cluster for holon formation.

Section 5 shows the how the concepts dealt with in the above articles align with this paper.

This theorem shows that holon formation is always possible. As shown later, non-holonic structures tend to disappear when the number of agents grows.

2.3 Holons and holonic information fusion

Information fusion (IF) is regarded here as a process of combining data from different sources to gain information that is more significant and useful than the unfused information gained from each source individually [20]. The IF process is typically divided into several levels [42]:

- data level, which combines data from different (essentially homogeneous) sources to produce a new, more useful and informative dataset than others.
- intermediate level, which combines previously gathered data to output a "features map" used to distinguish higher-level objects and relations.
- high level, which combines the decisions of several individual experts using voting, fuzzy logic and statistical methods for this purpose.

Likewise, holonic information fusion can simply be regarded as information fusion performed by means of a holon.

2.4 Real-world problems requiring a holonic approach with uncertainty. Resource-constrained networks.

While the communication possibilities between people and things have increased to the extent that, nowadays, almost everything communicates with almost everything, there are a number of factors (limited energy, communication costs, tactical risk of a high number of communications, limited computational power, limited bandwidth, etc.) preventing these networks from operating "normally" at all times (without constraints on the number of messages). Examples of this are: a) a network using mobile devices (for example, cell phones and tablets) with limited battery life or credit (applicable to communications between both people and things); b) networks subject to strict response time limits: each communication consumes time, which is sometimes a critical factor (for example, for handling incidents caused by natural disasters or acts of terrorism), and c) military networks with a high component turnover or edge networks [3,4].

Generically, networks with such constraints are called resource-constrained networks. They commonly include *ad hoc* wireless networks [35], mobile *ad hoc* networks (MANETs) [48,6,36,1], wireless sensor networks [2,23,56], delay-tolerant networks (DTN) [12,25,1], vehicular ad hoc networks (VANets) [26,11,36,1], and body area networks [15]. Specific applications of the above networks are distributed decision-making, real-time recommendation systems with geolocation, etc. The aim of the above networks is to achieve their goals by making optimal use of limited resources.

2.5 Holon formation

Several authors have studied the formation of holarchies under different circumstances. [50] demonstrates that when the holarchy is associated with the implementation of a plan that can be divided into subplans whose task definitions may overlap (that is, whose definition is vague), a holarchy minimizes the fuzzy entropy of the set of agents.

In [41] is set out the necessary and sufficient conditions for holon formation under a particular circumstance: it is necessary for the holon to have a greater minimum utility [56] its member holons and sufficient for it to have a greater maximum utility than any individual. Taking into account uncertainty, it can be said that it is necessary for the expected minimum utility to greater than the possible individual minimum utility and it is sufficient for the expected minimum to be greater than the possible individual maximum utility. These conditions can be regarded as a generalization to the non-empty "core" criterion in a cooperative game [5,16,14]: the holon emerges when all of its agents have a non-empty core as its action is regarded as cooperative play. This game theory-based approach is well suited for studying self-organization and possible "natural" associations generated between agents, as it studies strategic decision making in scenarios where there may be partnership agreements, with agents thus acting collectively [13][36] at the same time as competition: agents cooperate to achieve the global goal but at the same time compete for the communication capabilities. The model presented here illustrates this point. Andrade et al. [5] use the above approach to study the choice of partners in a holarchy. Besides, cooperative games with the characteristic function $v$, where $v(C)$ is the amount that the coalition $C$ could earn if its members decide to cooperate will be the most interesting as far as we are concerned here.

From the structural viewpoint, [31] singles out three types of relations that should necessarily occur in a system if it is to be classed as a holarchy: first-, second- and third-order relations. First-order relations exist within each holon (between structure and function or between data and process types). Second-order relations are associations of interdependency between holons, whereas third-order relations are associations occurring between holons of any level and the system as a whole, which are materialize as system coordination.

2.6 Self-organization

While we demonstrate that a system based on this model generates holons, they are formed unpredictably (their form cannot be determined a priori) because of the randomness at several levels regarding which data (fields) the agent will query, which agents will respond in time, agent response times, etc. On this ground, they can be regarded as the result of a process self-organization. The question of whether self-organization is feasible under these conditions is open. It is hard to define strict criteria that can be rigorously applied to decide whether or not there is self-organization. [33] classify the features of self-organization into two types: basic and autonomy-specific characteristics. A system does not need to have all these characteristics at once for there to be self-organization. For example, [38] takes no precautions against server overload, DoS (denial of service) attacks or the injection of corrupt data. Similarly, Gnutella [21] is devoid of both randomness and self-organized criticality, and they are, nevertheless, totally valid systems. The self-organization characteristics (taken from [34]) are detailed below.

*2.6.1 Basic characteristics*

1. Bounds: the system determines its own boundaries and should decide on new component membership.
2. Reproduction: the system can and should reproduce its structure (through addition, removal or change of a peer, its data, or relations or connections with other peers).
3. Mutability: the system must be capable of changing its structure, for example, changing the connections or forming clusters.
4. Organization: the system is organized either as a hierarchy, heterarchy or both.

5. Metrics: the system is capable detecting disturbances triggered by the environment, for example, failure of a peer or connection -in this case, when the other peer can be regarded as a failure-, data overload or manipulation (sending of "erroneous data").
6. Adaptability: the system is capable of properly reacting to disturbances (restructuring peers, adopting redundancy, etc.).

*2.6.2 Autonomy-specific characteristics*

1. Feedback: the system includes messages that peers send to each other. A P2P system is usually exposed to positive and negative feedback, on which ground system structure changes are balanced.
2. Complexity reduction: the system develops structures and hides details of the environment to reduce overall complexity (for example, through cluster formation or the creation of holons or other entities like active virtual peers).
3. Randomness: the system uses randomness as a prerequisite for creativity in order to effortlessly build a new structure.
4. Critically self-organized: a self-organizing system leads to a state of criticality. Appropriate procedures should be adopted to prevent too much order, as well as too much disorder, leading to greater flexibility, as the system can handle different types of disturbances. Too much order would be a hierarchical structure, and too much disorder a pure P2P.
5. Emergence: the system exhibits properties that are not characteristic of any peer or that were possibly unknown at design time.

*2.6.3 Additional characteristics*

Other possible characteristics of self-organizing systems are:

1. Peer equality counteracts the risk of single failure points.
2. A P2P system should ideally be capable of reproducing both its structure and its data. This should occur automatically and should not be triggered manually or come from outside.
3. Self-organization goes beyond the desirable properties of flexibility, adaptability and robustness. It includes the use of random components enabling the system to create viable new structures. In any case, the external control is confined to the minimum.

## 3 Problem statement

According to [31], a system has to carry out many more communications to perform a function if the control structure is centralized than if it is holonic: "The reduction in data transmission and in data complexity, achieved by the holonic architecture, is prodigious" [31]. Likewise, we want to investigate which structure emerges if the communication possibilities are reduced in a system that is the complete opposite of a centralized system, like, for example, an unstructured and decentralized P2P. Therefore, the problem can be described as follows:

> ***Given an open P2P system without a predefined structure, composed of agents (artificial or otherwise) behaving as described in Section 4 and carrying out IF (using data local to each agent) in order to increase the quality of the information contained in the messages that agents receive, holarchies are the predominant structure (topology) formed when system component communication resources (number of messages that they can send, message response time) are constrained and the components aim to be efficient.***

## 4 General properties of a system that implements the model

4.1 General characteristics of the model

Paggi, Soriano and Lara [40] describe a multiagent system model that aims to reduce the drawbacks of managing 'imperfect' information (with vagueness, uncertainty, incompleteness, etc.), using a ad hoc quality metric and information fusion. It is referred to as a system model because it provides neither a specification of how to perform function nor the vagueness or uncertainty metrics to be used.

The system responds to the message sent by the external agent (say $\Omega$). Each message is assumed to be composed of a series of components, about which a given agent can be more or less certain or which can be regarded as more or less vague (generically, as having higher or lower information quality) with regard to the querying agent. In turn, a component can be composed of others, etc. The messages can overlap: one can arrive before the previous one is answered.

Agents always aim to respond to the queries, and if they determine that their response is of lower quality than the query they may decide not to respond (there is a high probability of non-response, although there is a possibility, as shown later).

Probabilistic decisions are used in abundance, for example, when another agent is queried about a field $x$ with probability $1-quality(x)$ or when an agent can respond even though it is of lower quality than the querying agent with a probability that varies over time (an idea based on simulated annealing). This boosts self-organization and prevents local quality minimums. In turn, the model was designed such that self-organization leads to holonic structures.

Time is considered to be discrete, and increments (time steps) are as small as necessary.

The model assumes that all connections are possible, that is, that the network is an open "flat" (decentralized and unstructured [53]) peer-to-peer system, which, despite considering the number of time-outs, takes into account neither the communications quality nor cost, except the number of messages (occupied bandwidth could be used instead of this number, etc.) at any time. Whether or not an agent is considered to be physically local to another is not taken into account to calculate the cost. The resulting topology can be referred to as 'e-P2P' (and classed as evolutionary). It differences from the original flat P2P are shown in Table 1.

Table 1 - Differences between the two systems

| Pure P2P | e-P2P (holonic) |
|---|---|
| – All agents are queried and there are no favorite agents<br>– The querying agent waits for each queried agent<br>– There are no limits on the number of messages, where loops are limited by checking that the "parent" is not queried again. The number of visited nodes is not a practical anti-loop measure, as the metric would depend on a global number (total nodes), which, besides, is not constant due to peers entering and leaving<br>– The metric does not take into account whether the agent is of better quality than the querying agent at response time | – First all and then favorite agents are queried<br>– The querying agent waits for each queried agent, even if there is a time-out<br>– The number of messages is limited (there is no loop control, except checking the "parent" is queried again)<br>– Save in exceptional cases, queries are answered only if agent is of better quality than the querying agent |

The response time of each message is regarded as random for each agent and each interaction.

While all communications are possible, each agent selects who to query in the event of doubt (addressing its messages only to the agents that have recently responded with higher quality data).

As regards actual system operation, there are several agent phases:

- PHASE 0: An initial phase where the different agents are trained in the task to be performed.
- PHASE 1: A phase in which the agent determines which peers provide higher quality responses to the queries in each field. To do this, each agent sends the query to all the other peers ( or a set $w$ of peers, with a 'large enough' $w$ ) and then, after a specified number of interactions, determines which are the $p$ peers that have provided better quality responses, with $p \ll w$. This is the agent that then enters PHASE 2.
- PHASE 2: A phase where each agent exclusively queries its favorites, possibly fewer than $p$, and, when a peer does not respond (or does not respond in time), it tries with the next one on the list of peers with better quality performance.

All agents are assumed to pass through PHASE 0 together, although they may go back, and each agent passes through the different phases over time. When all the agents are in the same phase, the system is said to be in the respective phase.

Since, as time passes, the agents exhaust their communication capacity (which represents the battery life of a communications device, the messages credit of a prepay telephone, the physical possibility of making a call due to a hostile environment, etc.) and will stop responding, other agents that had not been queried because either they did not provide better quality responses, took too long respond or entered the system after the query was made are selected. This process resembles "natural selection": when an agent observes that too many time-outs are occurring, it simply switches to "intelligent" (or "economy") operating mode and starts to address messages exclusively to agents who have recently provided responses with higher quality data. Although new agents will join the system with time, the connection rate is assumed not to be as high as to be a clear case of churn [46]. Two non-exclusive approaches can be adopted to design a system that carries out decentralized IF: competitive or cooperative agents. In the competitive case, agents compete to provide independent measures of the same object property or characteristic. This can be very useful for critical systems (in the field of aerospace, for example) and fault-tolerant systems. On the other hand, the information provided by agents in cooperative environments is not redundant. The proposed system architecture is, on one hand, competitive, as each agent can, in theory, interpret (or, considering what it can deduce from other data, predict) all the compound fields, which it will try to do with better quality than the others, and, on the other, cooperative as this procedure is carried out such that final response of the whole is of maximum quality, and all the agents are benevolent. "Quality" is measured by a metric such that it is monotone decreasing with the error (uncertainty) and vagueness. For more details on this and other issues of the model, see [40] .

4.2 Self-organization and system autonomy

*4.2.1 System self-organization characterization*

The system implements the self-organization explained in Section 2.6 as follows:

*Basic characteristics*

- Limits: *In the proposed architecture, an agent that has a number of messages below a specified threshold does not recognize a new agent CHECK IN. In future versions, the new agent could also be asked to provide some certificate to attest to its identity or authenticity.*
- Reproduction: *Only logical connections, that is, with whom an agent communicates, change, as data replication makes no sense.*

- Mutability: *Mutability is the change in the above connections or the switch from unconstrained to intelligent mode.*
- Organization: the system is organized as either a hierarchy, heterarchy or both. *In this case, it is both (holons and other structures).*
- Metrics: *Although not accounted for in the current version, reputation learning and use could be used to manage false data cases, for example, through an adaptive mechanism that attaches more weight to peer that have a better reputation. Feedback reporting the real outcome would be required to calculate reputation and score the peer that comes closer to the real outcome higher. Additionally, the CHECK IN protocol could, in the future, include answering a query with a value already known to the bootstrap to evaluate whether the error specified by the agent that is attempting to enter is true (since this is one of the most important data that a peer can provide), where the new agent would not be able to query any other agent to get the specified control response (that is, the response given will be based exclusively on its skills).*
- Adaptability: *The adaptation is carried out by selecting new and better collaborators.*

*The system autonomy characteristics are:*

1. *The quality of the responses given by the components and the decision whether or not to respond are cases in point.*
2. *Most of the formed structures are holons, although other structures can be formed, as discussed in Section 5.*
3. *Randomness is used in different parts of the system: choices based on simulated annealing (when it decides whether to answer despite having a lower quality response than the querying agent) or the choice of fields and whether or not to query by such fields, etc.*
4. *The failure of any connections lead to a whole series of new query paths.*
5. *The system output a better quality result than any of the components separately, taking a relatively large number of system agents and a large number of messages.*

Other additional system characteristics, as mentioned in Section 2.7, are:

1. *All peers here are equal with respect to their system functions, except as regards the training that they have received.*
2. *The system automatically detects delays in agent responses and agents start to operate in intelligent mode).*
3. *The only external "control" event is the instruction given to agents that they have a number of available messages as of a particular time.*

4.3 Model applicability

We recommend the use of this model in situations where the following circumstances occur simultaneously (see [40]):

1. When the querying agent $\Omega$ repeatedly queries at a greater rate than peers enter and leave the system (that is, system composition is relatively stable) such that it makes sense to maintain a list of favorite peers.
2. When the number of favorite peers queried simultaneously is relatively small with respect to the total number of active peers in the system (when the opinion of a few is sufficient).
3. When the time-out time is long enough for peers to complete their processes (otherwise the peers leave due to time-outs and it is better not to use this model).

On the other hand, there are situations where the proposed model is not recommended for application:

1. When Condition 1 above is not met (as it would always end up asking all peers).
2. When the processing load of each peer cannot be increased any further (this model assumes that each peer requires extra CPU and memory usage to select and store its favorites). The extra usage will depend on the number of peers and the number of fields (for example, if there are 5000 agents in the system and 50 calculated (queryable) fields, the system would have to record the success rates of 250,000 combinations.
3. When it is important for a peer to retain its reputation after having left and re-entered the system. If a peer leaves the system (stops replying), it has to start to build its reputation from scratch when it re-enters (it has to start to generate good quality replies again from scratch).

## 5 Correspondence between holon concepts and formation

We show the correspondence between the concepts handled in the articles mentioned in Section 2 and this paper below. The fact that there is in both cases (this is, the mentioned articles and this work) an operation for clustering agents with a holon demonstrates that the system reported by [40] generates holons. Additionally, we show that the system dynamics lead to holon formation. We use the "▶" notation to abbreviate "equivalent due to the construction made with:

1. $\mathcal{A}$: Set of $n$ agents ▶ Set of $N$ peers $\{\alpha_1, ... \alpha_N\}$
2. $\varepsilon$ : Set of states in the system environment ▶ State of the environment of the peer $\alpha_i$, that is, $\varepsilon_i$ is defined by:
   - Limited messages indicator
   - Agent wanting to enter though $\alpha_i$ (CHECK.IN)
   - $\Omega$ waiting for response
   - the query message(s) sent to $\alpha_i$ by agents $\beta_1, \beta_2, ....$
   - the response(s) sent to $\alpha_i$ by agents $\beta_1, \beta_2, ....$

   The state of the system environment, $\varepsilon_S$ , is denoted (supposing that $\Omega$ does not belong to the system) by:
   - Limited messages indicator
   - Agent wanting to enter the system (CHECK.IN)
   - Query message(s) sent by $\Omega$ waiting for response

   – Response(s) sent to $\Omega$

   Therefore, $\varepsilon \equiv \varepsilon_S$
3. $\Pi$: perception function of agents $\Pi : \varepsilon \to P_1 \times ... \times P_n$ ▶ what the different agents perceive as the system environment
4. $S_i$ : agent state $\alpha_i$ ▶ state $S_i$ defined by
   – Number of available messages
   – Number of time-outs by queried agent
   – Intelligent or simple mode
   – Unanswered queries from other agents and their senders
   – Unanswered agent queries and their recipients
5. $\Phi_i$ : agent function $\alpha_i$ $\Phi_i : S_i \times P_i \to S_i \times A_i$ ▶ function agent of $\alpha_i$ (such as activity or operation) means that it switches from a specified number of available messages, number of queries to be answered or number of queries to be sent and a specified operating mode to another number of messages, queries, etc., after performing a particular action.
6. $P_i$ : agent $\alpha_i$ perceptions ▶ possible agent perceptions are:
   – Messages received (queries and responses) and their senders
   – TIME OUTS and their originators
   – CHECK IN phase requests
   – Specification of limited messages
7. $A_i$: set of agent $\alpha_i$ actions ▶ possible system agent actions (in terms of how it affects its interaction with others) are:
   – Answer query
   – Ignore query
   – Make query
   – Answer CHECK IN request
   – Ignore CHECK IN request
   – Switch to intelligent mode
8. $\bar{\Delta}$: change of system state $\bar{\Delta} : \varepsilon \times S_1 \times ... \times S_n \to \varepsilon \times S_1 \times ... \times S_n$ ▶ system state changes whenever either the state of its environment or any of its agents changes.
9. $\Pi'$ : perception function with a formed holon. Construction of $\Pi'$: it is constructed using a vector of perceptions as the holon perception, as shown in Section 2.2. ▶
   – Let $\alpha^*$ be a holon formed based on agents $\{\alpha_1, ...\alpha_k\}$:
   – When an agent determines its $k$ best peers for any field, a holon ($\alpha^*$) is being formed whose head is the agent and whose body is the $k$ peers.
   – Agents can participate in more than one holon: an agent can be one of the best in a field $X$ for one query and one of the best in another field $Y$ for another
   – $\Omega$ is also assumed to choose the best agents for its query

   Formally:
   **Given a set of agents** $\{\alpha_1, ...\alpha_k, \alpha_{k+1}, ...\alpha_N\}$
   – **The original cluster $H$ of $\{\alpha_1, ...\alpha_k\}$ takes place when $\Omega$ determines the best $k$ agents for its query. The head of $H$ will be $\Omega$ and the $\{\alpha_1, ...\alpha_k\}$ heads of its subholons. The above holon is autonomous and communicates by means of its head.**
   – **Given the pair $\{H, \alpha\}$, $H$' is formed when any agent $\beta$ of any $H_i \subseteq H$ determines that $\alpha$ is one of the $k$ best agents (of the best sources) for any query type. The head will be $\beta$ and) $\alpha$ and the agents that it queries (favorites) plus any other subholons already in $\beta$. The probability of the communication not taking place through $\beta$ decreases with $N$ (see Figure 2).**
   – **Given $\{H_1, H_2\}$, $H$' is formed when any agent $\beta$ that does not belong to either $H_1$ or $H_2$ determines that the heads of $H_1$ and $H_2$ are among the best $k$ sources for any query type. The head of $H$' will be $\beta$ and $H_1$ and $H_2$ will be part of its body.**
10. $\bar{\Delta}'$: Change of system state with one defined holon ▶ It is constructed as described in Section 2.2.

11. Abstract resource: Any device (including information) or a tool the improves agent behavior▶ The resources managed by agents are:
    – Energy (battery) – not actually modeled
    – Available messages
    – Predictions of other agents and qualities
    – Limited messages indicator
12. $u$: Utility function. $u : E \times S \to \mathbb{R}$. For the $i$-th agent, $u_i : E_i \times S_i \to \mathbb{R}$▶ the resulting utility for agent $\alpha_i$ is measured as $Q$ : the information quality that the system can offer (the querying agent) in return
13. $E_i$: agent environment $\alpha_i$▶$E_i$ includes the other agents and $\Omega$
    – The set of its states is $\varepsilon_i$ defined previously
14. Resource allocation mechanism:
    – It is controlled by the holon members through individual negotiation
    
    Market-based mechanism:
    Each agent publishes a task or resource and the other agents place bids for the resource or perform the task to increase the quality of the distribution.
    It is found that
    – Allocation is not cooperative, as each agent tries to maximize its local quality without taking into account the global result
    – Agents are altruistic (they work without reward)
    
    It is equivalent by construction to:
    – Resources are allocated when agent $\alpha_i$ selects the best collaborators for a particular field. Technically, it is a sealed-bid-first-price auction (the highest bid wins, irrespective of who placed the bid).
    
    Finally, the hypothesis stated in Section 2.2 holds in this case as it is a flat P2P system.

Taking into account the phasewise operation explained in Section 4.1 and considering how the holons are formed according to item 9 above (highlighted in bold), it is clear that the system generates holons during operation.

While the above formed structures may not communicate exclusively through their heads (see Figure 2), this becomes extremely unlikely as $N$ grows; that is, the probability of not being a holon tends to 0. This is evident considering the individual probabilities of there being a situation, such as is shown in Figure 2. Supposing that:

– the agents make queries independently (that is, one agent's query does not influence another's),
– an agent cannot query the agent that is querying it,
– neither the percentage of available resources for each peer at any time (probabilities are lower if this is taken into account) nor the quality of the message that are being communicated are taken into account such that the communication is possible

we find that:

1. P($\gamma$ is chosen as the favorite of $\alpha$)=$1/(N-1)^C$
2. P(K given agents are the favorites of $\alpha$)=$1/(N-1)^{CK}$
3. P(situations 1 and 2 above hold for agents $\alpha, \beta, \gamma$ )=$p$

$$p = \left( \frac{1}{N-3} \frac{1}{N-2} \frac{1}{N-1} \right)^{CK} \quad (\textbf{Eq. 1})$$

4. P(situation 3 applies for any tuple of system agents)= $p'$

$$p' = A_3^N \left( \frac{1}{N-3} \frac{1}{N-2} \frac{1}{N-1} \right)^{CK} = \frac{N(N-1)(N-2)}{((N-3)(N-2)(N-1))^{CK}} \leq \frac{N(N-1)(N-2)}{(N-3)^{3CK}} \quad \textbf{(Eq. 2)}$$

where

- $K$ is a system parameter = number of agents to be taken into account as agent "favorites",
- $C$ is another parameter: the number of interactions that take place before an agent is chosen as favorite (PREVIOUS.MESS),
- $P(\bullet)$ denotes probability, as usual.

The probability $p'$ is regarded as very small for relatively low values of $C$, $K$ and $N$. Additionally, this probability is calculated assuming that agents have access to the resources that they need to perform all the necessary communications, on which ground the final probability $p''$ is even smaller:

$$p'' \leq \frac{N(N-1)(N-2)}{((N-3)(N-2)(N-1))^{CK}} \leq \frac{N(N-1)(N-2)}{(N-3)^{3CK}} \approx \frac{1}{N^{3CK-3}} \quad \textbf{(Eq. 3)}$$

For example, $p'' \leq p' \approx 1/20^{3*3*5-3} = 1/20^{42} = 0.5E-42$ for 20 agents, three interactions and five favorites, which is negligible.

The values $p'$ and $p''$ indicate the probability of agent groups not communicating exclusively through the head, that is, of their not being holons. As such probability drops rapidly as $N$ grows, we conclude that all the structures are holons, that is, P (the structure formed is a holon) = 1 when $N \to \infty$.

The theorem explained in Section 2 assures that a holon can be generated as long as these values show that they are all holons on the edge.

## 6 Example

An example designed to clarify the operation of a model implementation is described below.

The tasks to be performed are classification, that is, decision-making support. The aim is to classify whether or not an automobile is acceptable based on a set of attributes. This example builds on a case inspired in a data set related with the classification of a car as appropriate or not for buying, depending on a series of attributes such as model, brand, sale price, etc. similarly as in the data set "cars" in [28]. Our work does not compare the model proposed with any other but instead it just explains some of its properties.

6.1 System dynamics

Let $\Omega$ be the sender agent that sends a message. Let $\alpha, \beta, \gamma$ be the set of agents of the example. The maximum permitted number of messages per agent is 10.

The registers to be sent have fields A,B,C,D such that A is formed based on B, C, C from D, D from E, F, G, and B depends on G: $A = f(B,C)$, $C = g(D)$, $D = h(E,F,G)$ and $B = p(G)$ where $f, g \ldots p$ are specific functions.

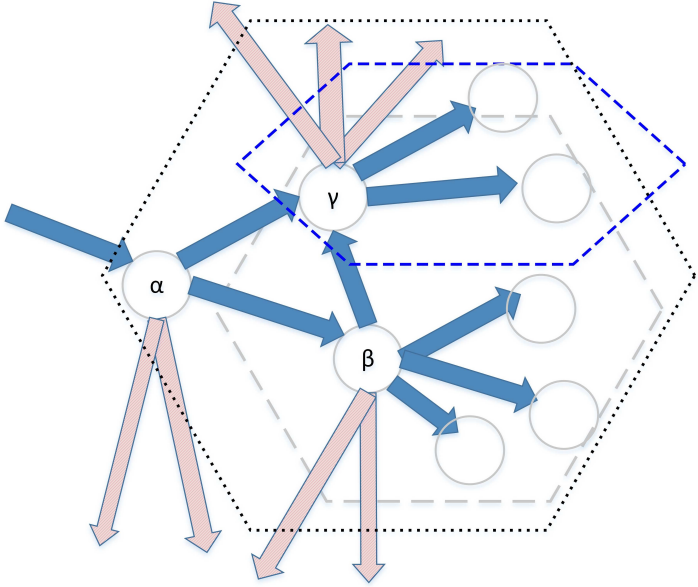

**Fig. 2** Case of abnormal communication (The dark-colored arrows denote the communications with favorites; the hexagons denote holons)

Let M be the complete message sent by $\Omega$ or let M=A||B||C||D||...||G where || represents the concatenation as usual. Fields A...G maps with car attributes. In particular, let M be the data tuples related to the comfort of different vehicle types and prices:

- A= comfort rating (integer from 1 to 5, where 5 is the maximum)
- B= no. of doors (2 to 5)
- C= equipped with air conditioning (Y/N)
- D= sale price (in US dollars)
- E= vehicle type (utilitarian: Y/N)
- F= vehicle source (national or imported, where 1 = national)
- G= model (sedan, coupé, etc.)

All the fields in the final dataset were converted to numerical fields.

All the data are assumed to be "crisp", that is, unambiguous. An agent can only have uncertainty with regard to compound fields (that is, fields that depend on another or others; we do not account for any ´error for fields that are not compound or dependent). This uncertainty is the agent's "prediction" error for that field. The prediction error for message M then is to wrongly classify the record "decide whether a vehicle provides enough comfort for its characteristics"(binary decision).

The prediction errors (as a percentage)for each compound field and each agent output after the training stage are shown in Table 2, that should be taken to mean, for example, that the error made by agent α when predicting A is 5%.

Table 2. Errors of the peers

|   | A   | B   | C   | D   | M   |
|---|-----|-----|-----|-----|-----|
| α | 5%  | 10% | 15% | 20% | 5%  |
| β | 25% | 10% | 10% | 15% | 10% |
| γ | 10% | 15% | 20% | 15% | 25% |

There is a stabilization period during which $\Omega$ sends a message to all the agents, after which the agent decides to only send messages to that agents that have answered more often with the lowest error. The same applies to all the other agents.

Each agent presumably decides to send the message to the agent that has provided the best response most often, for which purpose it waits until it has received only one response (that is, it sends only one message to find out which is the most successful agent rather than sending, for example, 20 messages and then examining the success rates of each agent to select the agent with the highest success rate). For simplicity's sake, we also measured the quality of a field that is inversely proportional to the prediction error of the above field without considering the other variables (number of query forwards, number of queried agents, etc.). etc.).

When an agent intends to query a peer about field X, it queries all the agents that it has previously queried about X that have a number of successes equal to the maximum. Each agent (including $\Omega$ ) is assumed for this purpose to have a table containing the above rates ($\Omega$ only of M): the "BEST-0" table (Table 3) for $\Omega$ includes the number of times that each agent returns the best response and BEST (Table 4) the number of times an agent gave the best answer for a field, for every agent.

Table 3. "BEST-0"

| Field | α | β | γ |
|-------|---|---|---|
| M     | 0 | 0 | 0 |

BEST-0($\beta$)=3 would mean that $\beta$ is the agent that has provided the best response to message about M sent by $\Omega$ so far.

Note that M is the entire message M=A||B||C||D||E||F||G rather than just a field. M= comfort||no. doors||air conditioning||price||...

All the table fields are set to 0 at the start, that is, before any message is received.

The number of remaining messages are also counted (see Table 5).

Table 4. "BEST"

| For $\alpha$ | $\alpha$ | $\beta$ | $\gamma$ | For $\beta$ | $\alpha$ | $\beta$ | $\gamma$ | For $\gamma$ | $\alpha$ | $\beta$ | $\gamma$ |
|---|---|---|---|---|---|---|---|---|---|---|---|
| A | | | | A | | | | A | | | |
| B | | | | B | | | | B | | | |
| C | | | | C | | | | C | | | |
| D | | | | D | | | | D | | | |
| M | | | | M | | | | M | | | |

Table 5. "REMAINING-MESSAGES"

| Agent | Number of messages |
|---|---|
| $\alpha$ | 10 |
| $\beta$ | 10 |
| $\gamma$ | 10 |

that is, there are 10 remaining messages for $\alpha$ at this time.

Each agent has a probability proportional to its prediction error for a given field of querying another agent about the respective field (that is, if error is high, uncertainty is high and there is a high probability that the agent will launch a query). Note that prediction error is determined at training time (PHASE 0).

In this example, the fusion of all the possible responses received by an agent involves the choice of the response with the least error.

The mentioned tables evolve over time: the table of the number of times that each agent provides the best response to the "field" M for $\Omega$ evolves as shown in Table 6.

Meaning that, for instance, at time $t = 25$, $\alpha$ had responded three times to the query by $\Omega$, whereas the others had not responded even once, on which ground $\alpha$ is the best agent (at time 25) according to $\Omega$.

For other agents, the evolution is shown in Table 7.

Leading to changes at specified times (such as at $t = xx$), that is, when $t = 18$ for $\alpha$ the agent that best responds to its query about field C was $\beta$.

The number of remaining messages per agent evolves over time according to Table 8.

For instance, at time 25, agent α has three remaining messages, whereas β has four.

Figure 3 shows a diagram based on AUML [39] that illustrates system evolution. The vertical axis represents the time step whereas the vertical bands represent the different agents, the solid arrows symbolize query messages and the dashed arrows query responses. Each agent knows the number of remaining messages that it has but not those that others have. Agent response time is random.

Figure 3 details are explained as follows:

• The dashed arrows represent responses, the solid arrows, requests. The vertical lines mean that a message generated other query messages.

Table 6 - Evolution of "BEST-0" table

| $t$ | $\alpha$ | $\beta$ | $\gamma$ | $t$ | $\alpha$ | $\beta$ | $\gamma$ |
|---|---|---|---|---|---|---|---|
| 1 | 0 | 0 | 0 | 26 | 3 | 0 | 0 |
| 2 | 0 | 0 | 0 | 27 | 3 | 0 | 0 |
| 3 | 0 | 0 | 0 | 28 | 3 | 0 | 0 |
| 4 | 0 | 0 | 0 | 29 | 3 | 0 | 0 |
| 5 | 0 | 0 | 0 | 30 | 4 | 0 | 0 |
| 6 | 0 | 0 | 0 | 31 | 4 | 0 | 0 |
| 7 | 0 | 0 | 0 | 32 | 5 | 0 | 0 |
| 8 | 0 | 0 | 0 | 33 | 5 | 0 | 0 |
| 9 | 0 | 0 | 0 | 34 | 6 | 0 | 0 |
| 10 | 0 | 0 | 0 | 35 | 6 | 0 | 0 |
| 11 | 0 | 0 | 0 | 36 | 6 | 0 | 0 |
| 12 | 0 | 0 | 0 | 37 | 6 | 0 | 0 |
| 13 | 0 | 0 | 0 | 38 | 6 | 0 | 0 |
| 14 | 1 | 0 | 0 | 39 | 6 | 0 | 0 |
| 15 | 1 | 0 | 0 | 40 | 6 | 0 | 0 |
| 16 | 1 | 0 | 0 | 41 | 6 | 0 | 0 |
| 17 | 1 | 0 | 0 | 42 | 6 | 0 | 0 |
| 18 | 1 | 0 | 0 | 43 | 6 | 0 | 0 |
| 19 | 1 | 0 | 0 | 44 | 6 | 0 | 0 |
| 20 | 2 | 0 | 0 | 45 | 6 | 0 | 0 |
| 21 | 2 | 0 | 0 | 46 | 6 | 1 | 0 |
| 22 | 2 | 0 | 0 | 47 | 6 | 1 | 0 |
| 23 | 3 | 0 | 0 | 48 | 6 | 1 | 0 |
| 24 | 3 | 0 | 0 | 49 | 6 | 1 | 0 |
| 25 | 3 | 0 | 0 | 50 | 6 | 1 | 1 |

Table 7 - Evolution of the "BEST" table

| For $\alpha$ | $\alpha$ | $\beta$ | $\gamma$ | For $\beta$ | $\alpha$ | $\beta$ | $\gamma$ | For $\gamma$ | $\alpha$ | $\beta$ | $\gamma$ |
|---|---|---|---|---|---|---|---|---|---|---|---|
| A | 1 (at t=6) | | | A | | | | A | | | |
| B | | | | B | | 1 (at t=9) | | B | | | |
| C | | 1 (at t=18) | | C | | | | C | | | |
| D | | | | D | | 1 (at t=10) | | D | | | |
| M | | | | M | | | | M | | | |

- P(Y,Z) means "prediction, made by Y, of field Z".
- The letters in the boxes refer to the queried field.
- Time elapses vertically, from top to bottom. Each row represents a time ($t$), numbered 1, 2...
- A Ø means that there is no response because it has a larger error. If there is no response because it has no messages left, then the response line is not represented.
- $M_i$ represent messages ( $M_i \neq M_j$ with $i \neq j$).
- A holon H whose head is an agent U is formed when U is the first option of whichever agent queries H. H is said to have emerged as of when U becomes the best option.

Table 8. Evolution of "REMAINING-MESSAGES"

| $t$ | $\alpha$ | $\beta$ | $\gamma$ | $t$ | $\alpha$ | $\beta$ | $\gamma$ |
|---|---|---|---|---|---|---|---|
| 1 | 10 | 10 | 10 | 26 | 2 | 4 | 10 |
| 2 | 10 | 10 | 10 | 27 | 2 | 3 | 10 |
| 3 | 10 | 10 | 10 | 28 | 2 | 2 | 10 |
| 4 | 10 | 10 | 10 | 29 | 2 | 2 | 10 |
| 5 | 9 | 10 | 10 | 30 | 2 | 2 | 10 |
| 6 | 8 | 8 | 10 | 31 | 2 | 2 | 10 |
| 7 | 8 | 6 | 10 | 32 | 1 | 2 | 10 |
| 8 | 8 | 6 | 10 | 33 | 1 | 2 | 10 |
| 9 | 8 | 5 | 10 | 34 | 0 | 2 | 10 |
| 10 | 8 | 5 | 10 | 35 | 0 | 2 | 10 |
| 11 | 8 | 5 | 10 | 36 | 0 | 2 | 10 |
| 12 | 8 | 5 | 10 | 37 | 0 | 2 | 10 |
| 13 | 7 | 5 | 10 | 38 | 0 | 2 | 10 |
| 14 | 7 | 5 | 10 | 39 | 0 | 0 | 10 |
| 15 | 7 | 5 | 10 | 40 | 0 | 0 | 9 |
| 16 | 6 | 5 | 10 | 41 | 0 | 0 | 9 |
| 17 | 5 | 5 | 10 | 42 | 0 | 0 | 8 |
| 18 | 5 | 4 | 10 | 43 | 0 | 0 | 8 |
| 19 | 5 | 4 | 10 | 44 | 0 | 0 | 8 |
| 20 | 4 | 4 | 10 | 45 | 0 | 0 | 7 |
| 21 | 4 | 4 | 10 | 46 | 0 | 0 | 7 |
| 22 | 4 | 4 | 10 | 47 | 0 | 0 | 7 |
| 23 | 3 | 4 | 10 | 48 | 0 | 0 | 7 |
| 24 | 3 | 4 | 10 | 49 | 0 | 0 | 7 |
| 25 | 3 | 4 | 10 | 50 | 0 | 0 | 7 |

- An agent answers the query only if it has a strictly lower prediction error than the querying agent, and if it has messages left.
- When $\Omega$ starts to operate, it waits until it has at least two responses in order to decide which is better or until there is a time-out.

Figure 4 shows the major milestones illustrated in Figure 3, where time elapses from top to bottom.

6.2 Holon formation

The diagram shown in Figure 5 shows holon evolution for the illustrated example. Figure 6 depicts another possible holons evolution when $N = 2$.

The marked times 14, 36, 46, 49 signal the time at which each phase starts and when the agent or holon $\Omega$ starts to send messages to a different agent. For example, at time 14, it switches to query $\alpha$ only and at 46 $\beta$ only. At 36, $\alpha$ no longer has any messages and does not respond, on which ground $\Omega$ sends messages to other agents until it manages to select a new agent. See the time evolution in the BEST table.

Each iteration has two well-defined phases: PHASE 1 and PHASE 2. In PHASE 1 of the first iteration, we find that all the agents are working as a

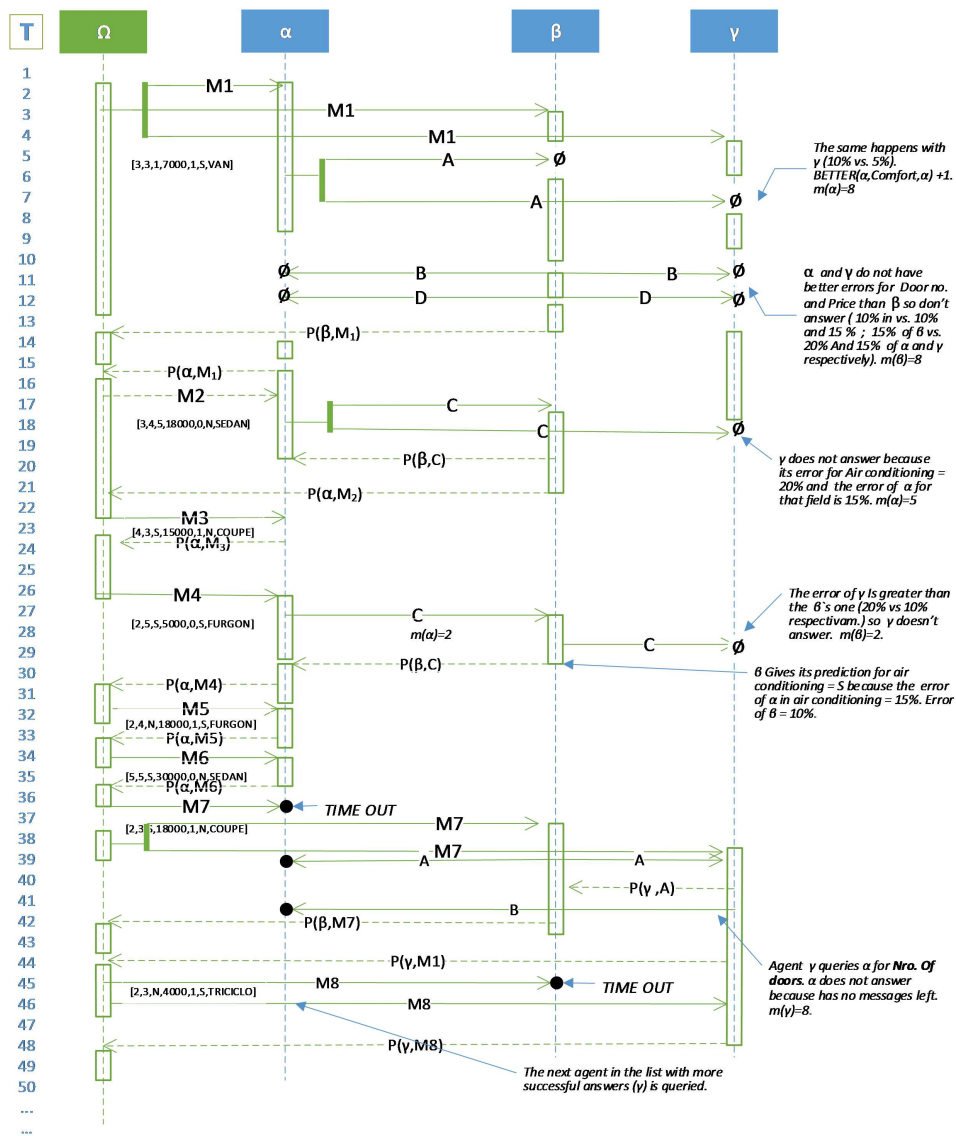

**Fig. 3** A possible time evolution of the system

P2P. After a stabilization period, $\Omega$ learns which agent to send messages to. This is PHASE 2, where $\Omega$ has the agent that provided the best response. A holon can then be said to have been generated. In this case, the holon emerges at time 14 and disappears at time 36. Intelligent system behavior emerges at times 14 and 46.

| T | Comments |
|---|---|
| 1 | α queries β and γ about the field Comfort. M(α)=10, m(β)=10, m(γ)=10 |
| 2 | β queries about Doors no. and Price |
| 4 | Agents have a random response time, so in this instant nothing happens. |
| 5 | β has greater error for Comfort than α (10% vs. 5%) so it does not answer. m(α)=9 |
| 6 | The same happens with γ (10% vs. 5%). BEST(α,Comfort,α) +1 m(α)=8 |
| 8 | α and γ don't have better errors for Doors no. and Price than β and don't answer ( 10% of β vs. 10% and 15 % ; 15% of β vs. 20% and 15% of α and γ respectively). m(β)=8 |
| 9 | m(β)=6. BEST(β, Doors no., β) +1. BEST(β, price, β) +1 |
| 11 | β gives its prediction of M altogether its quality (error): Y (Adequate). Error=10%. m(β)=5 |
| 13 | Idem α : answers N (Not Adequate). Error=5% . m(α)=7 |
| 14 | Ω will decide how to classify the vehicle from the answers. BEST-0(α)+1 (and reaches 1) |
| 15 | The following queries are addressed to α because it was the one with the best performance, while it can answer. |
| 16 | α queries about the field Air conditioning. m(α)=6 |
| 17 | γ does not answer because its error for Air conditioning = 20% and the error of α for that field is 15%. m(α)=5 |
| 18 | β gives its prediction for Air conditioning altogether its quality (error). Error= 10%. Air conditioning = Y. BETTERS(α, air conditioning, β) +1 |
| 20 | α gives its prediction for m: **Comfort = N (not adequate). Error = 5%.** BEST-0(α)+1 (and reaches 2). m(α)=4 |
| 23 | α gives its prediction for m: **Comfort = Y (Adequate). Error = 5%.** BEST-0(α)+1 (and reaches 3). m(α)=3 |
| 25 | α queries β who in turn queries γ about **Air conditioning** |
| 26 | m(α)=2 |
| 27 | m(β)=3 |
| 28 | The error of γ is greater than the β's one (20% vs 10% respectively) so γ does not answer. m(β)=2 |
| 29 | β gives its prediction about **Air conditioning** = Y because the error of α for **Air conditioning** = 15%. Error of β = 10% |
| 30 | Prediction of α for M4: **Comfort = N ( not adequate). Error = 5%.** BEST-0(α) +1 (and reaches 4) |
| 32 | Prediction of α for M5: **Comfort = Y (Adequate). Error = 5%.** BEST-0(α) +1 (queda en 5) |
| 34 | Prediction of α for M6: **Comfort = N (Not adequate). Error = 5%.** BEST-0(α) +1 (and reaches 6). m(α)=0 |
| 35 | From here on α has no messages left. The next in sequence of frequencies are queried (β y γ) |
| 37 | Query others with answer frequency lesser than α (β y γ) |
| *39 | β queries γ and α about **Comfort.** m(β)=0 |
| 40 | Error of γ for **Comfort** = 10%. Prediction of γ for **Confort**= 3 . m(γ)= 9 |
| 42 | Agent γ queries α about **Door no.** α does not answer because has no messages left. m(γ)=8 |
| 44 | β answers and runs out of messages. Prediction of β para $M_6$ **Comfort = Y (Adequate).** m(β)=4. **Error = 10%** |
| 45 | *This message is ignored by α for being out of time.* Agents take a random time to answer. Prediction of γ for $M_6$ **Comfort = Y. Error = 25%.** m(γ)=7 |
| 46 | BEST-0(β) +1 (and reaches 1) |
| 49 | The next agent in the list with more successful answers is queried (γ) |
| 50 | Prediction of γ for $M_8$ **Comfort = Y. Error = 25%.** BEST-0(γ) +1 (and reaches 1). m(γ)=6 |

**Fig. 4** Interesting milestones for the example

In SITUATION 1, the agent $\alpha$ can be regarded as acting as the holon head, operating as the group's query input/output gate and coordinating the results of the other two agents ($\beta$ and $\gamma$). In turn, $\beta$ and $\gamma$ operate as the body, processing other parts of the message. The same applies in SITUATION 3, where $\alpha$ is the head and $\gamma$ is the body.

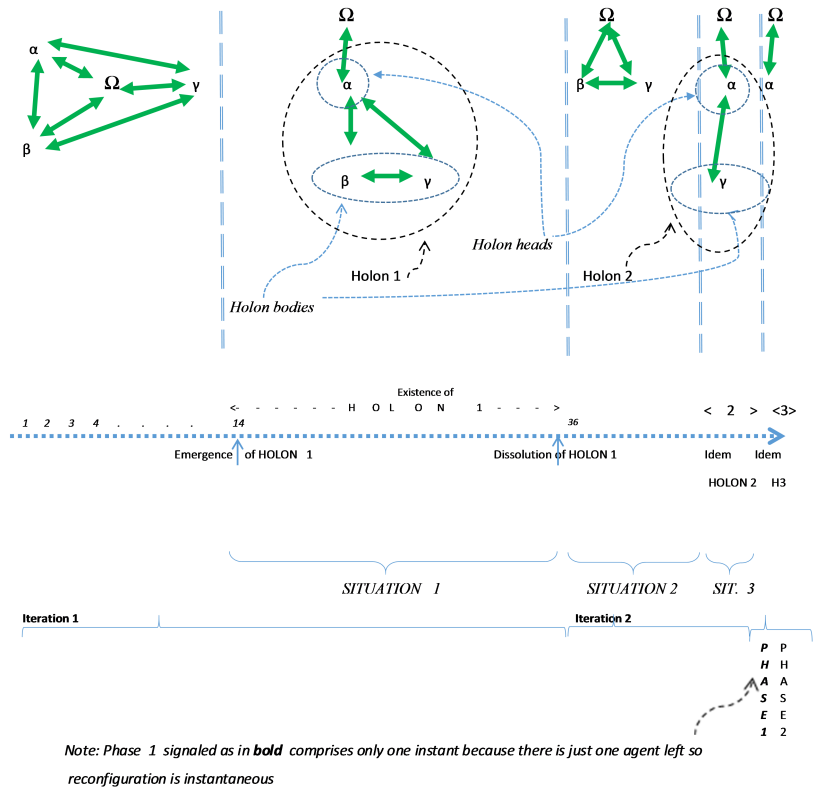

Fig. 5 Holon evolution

Note that the P2P and holonic structures alternate as the number of messages drops. They are holons because they are sets of semi-autonomous, possibly heterogeneous, agents, grouped by levels where a head-body structure is evident at each level, formed to achieve a common goal. They use level-specific information to do their jobs.

Likewise, the conditions to be met, according to [22], by a holarchy also hold:

1) Autonomy: agents that answer queries are autonomous and may or may not respond, although, when they participate in a holon, they lose some autonomy as they have to perform tasks for the holon for a time.

2) Goal-driven behavior: agents cooperate to answer an incoming query with the best possible quality information.

3) Extended group capabilities: holon response includes information to which its members do not have access separately (for example, number of agents that responded, differing qualities, etc.). Both response quality and numbers may be lower than for queries due to the formulas or the simulated annealing that it uses (see Section 4.1).

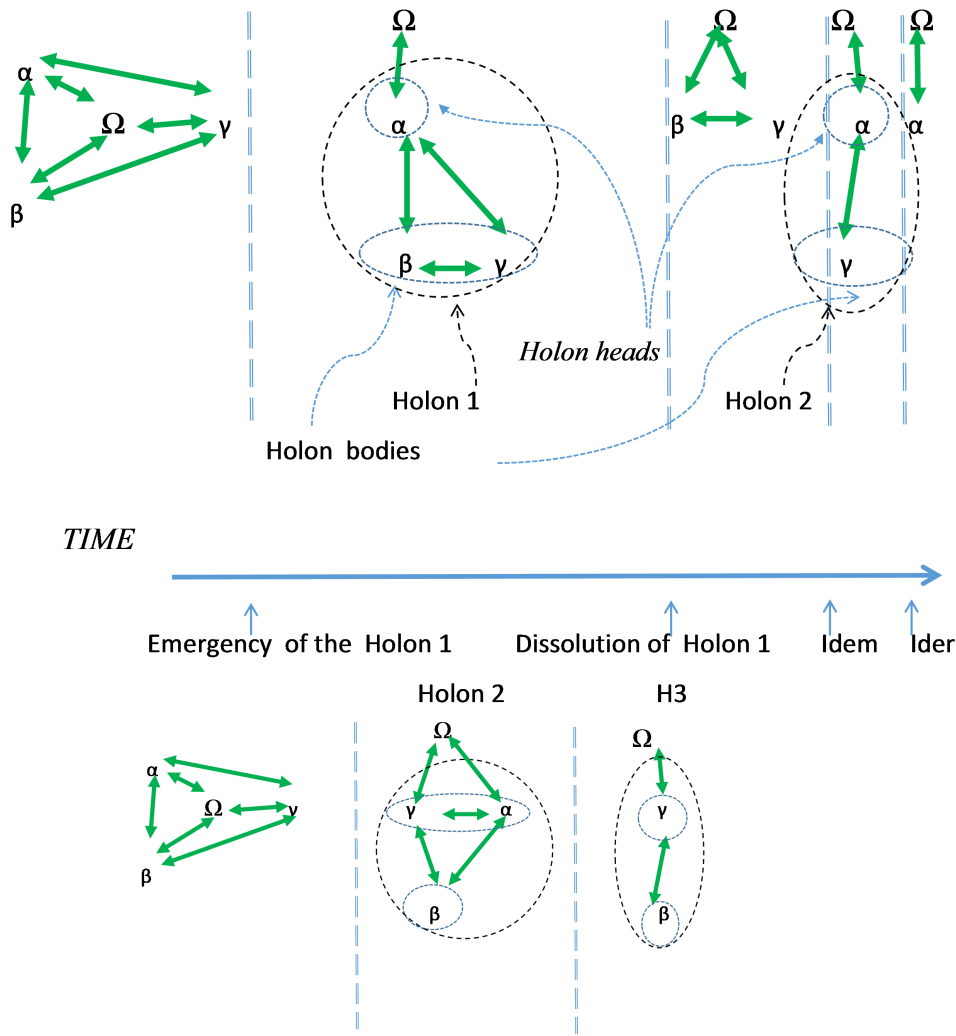

**Fig. 6** Possible holon evolution (The agents queried by $\Omega$ are $\alpha, \beta, \gamma$; At the top: PREVIOUS.MESS=1. They know the agent messages limit from the start. No agents enter the system during the studied period. N = 1; At the bottom: The situation is the same, save that N=2. Emergence takes place at other times)

4) Beliefs: the agents have explicit representations of the environment (preference tables, list of active agents, etc.).

5) Limited rationality: the super holon monitors the use of resources by the subholons. In this case, it does not directly control the use of resources, although it can be said to control them indirectly: if it has fewer than $x\%$ of messages, it draws lots for decide if it answers the query. In case of not querying, this prevents the subholon from wasting its resources on responding, as it is possible that it responds but the querying agent cannot finally do it because it has used up all its messages.

6) Communication: the communications between subholons necessarily pass through the head. When holons have been formed, the communications pass through the heads only, save in exceptional cases. This holon condition was discussed in Section 5.

## 7 Conclusions and future work

We showed the equivalence between the concepts and entities handled in the system and reported by [22] and thus were able to prove that a system implementing this model can generate holons and that, in the limit when $N \to \infty$, all the formed structures are holonic. Holon formation was predictable as the model was designed in order to boost self-organization and holon generation.

Holons have a series of advantages: they are semi-autonomous, are fault- and internal or external disruption-tolerant, can distribute control load, etc.

While it is known that holons will be generated, it is impossible to discover beforehand which holons they will be. Holons are therefore the result of agent self-organization. We show that they meet the conditions required for self-organization.

Systems of this type can be applied to handle large volumes of data, such as big data and data analytics, where each agent specializes and processes a particular data subset using specialized analysis tools. Another example could be a mobile recommender system. The query posed by an user is answered by different smart phones forming the P2P system which can use context information of that user as data fields (geoposition, etc.). Clearly as long every phone exhaust their batteries or messages they leave the system logically and another peer has to be queried but the network continues working smoothly because of the adaptation ability of the holons. The simulated annealing-style characteristics allows to foster the serendipity driving to try options that in other circumstances would be ignored or rejected.

Future lines of research are:

– Analyze whether feedback and the resulting agent learning should be enabled. An agent in this model never knows whether the response it gave was the best or was received too late (time-out). Adding related feedback from the sender (the querying agent) would increase the number of messages to be sent; however, it would trigger intelligent behavior on the part of the receiver at an earlier stage (the agent, finding that it is producing low quality results, will opt to switch to intelligent mode), reducing the number of messages (e.g., leading the receiver to query other agents to get a better quality response). This is, therefore, a design issue warranting evaluation.

- Enable a different time-out for each agent (and thus have "nervous" and "calm" agents).
- Analyze what is "relatively" as mentioned in Section 4.1
- Study what changes to make to the model when the agent is unclear about how to decompose a given field (or, alternatively, about other fields of which it is part). A topic worth researching is the choice of one out of several feasible decompositions. Generally, we have assumed that all agents share the decomposition criterion (vague or otherwise) of the messages and fields (e.g., by using the same ontology).
- Analyze how this model can improve the performance of sensor holarchies. The model can be useful for sensors that exchange information. In fact, studies have been conducted on information fusion from sensors [7].
- Study how this system model can contribute to some of the open problems in the big data area. As in the big data and big data processing are distributed across different repositories, it makes sense for the different model agents to specialize in processing different data subsets. The preliminary training (PHASE 0) could be carried out by specializing agents or, alternatively, agents that enter the system could be conceived as already specialized in a specified data subset or view.

**Conflict of Interest: The authors declare that they have no conflict of interest.**